\documentclass{article}
\usepackage[utf8]{inputenc}
\usepackage[T2A]{fontenc}
\usepackage[english]{babel}
\usepackage{amsmath}
\usepackage{amsfonts}
\usepackage{appendix}
\usepackage{graphicx}
\usepackage{float}
\usepackage{csquotes}

\usepackage{subcaption}
\usepackage{wrapfig}

\usepackage[parfill]{parskip}

\usepackage[
backend=biber,
style=numeric-comp,
sorting=none  
]{biblatex}
\usepackage{hyperref}
\title{Cluster Generation via Deep Energy-Based Model}
\author{
A.~Yu.~Artsukevich \\
artsukevich@gmail.com
\and 
S.~V.~Lepeshkin$^{1,2,3}$ \\
S.Lepeshkin@skoltech.ru
}

\date{%
    $^1$ Skolkovo Institute of Science and Technology, Skolkovo Innovation Center, 3 Nobel St., Moscow, 143026, Russian Federation\\%
    $^2$ Vernadsky Institute of Geochemistry and Analytical Chemistry, Russian Academy of Sciences, Kosygina, 19, Moscow, 119991, Russia\\
    $^3$ Lebedev Physical Institute, Russian Academy of Sciences, 53 Leninskii prosp., 119991, Moscow, Russian Federation\\%
}

\begin{document}

\maketitle

\begin{abstract}
We present a new approach for the generation of stable structures of nanoclusters using deep learning methods. Our method consists in constructing an artificial potential energy surface, with local minima corresponding to the most stable structures and which is much smoother than "real" potential in the intermediate regions of the configuration space. To build the surface, graph convolutional networks are used. The method can extrapolates the potential surface to cases of structures with larger number of atoms than was used in training. Thus, having a sufficient number of low-energy structures in the training set, the method allows to generate new candidates for the ground-state structures, including ones with larger number of atoms. We applied the approach to silica clusters $(SiO_2)_n$ and for the first time found the stable structures with $n=28\ldots 51$. The method is universal and does not depend on the atomic composition and number of atoms.

\end{abstract}

\hspace{10pt}

\section{Introduction}

The particles that are sized in order of several nanometers can be found among the most interesting research objects in different scientific domains like condensed matter physics, materials science and other related fields. Being settled in between molecules and bulk solids, these particles (nanoparticles, nanoclusters or quantum dots) exhibit many unique properties, contrasting to ones of bulk solids, related to the same material. These properties are closely related to atomic structure of the particles, especially in case of small particles and nanoclusters. Despite the importance of knowing the atomic structure, it’s experimental determination remains rather difficult. For this reason the main body of structural information about nanoclusters is gathered with first-principles calculations. 

Nowadays the widely acceptable approach to predicting the cluster structure is based on global optimization. A number of corresponding methods have been developed - basin \cite{basinhopping1997} and minima hopping \cite{goedecker2004minima}, simulated annealing \cite{wille1986searching, schon1996first}, metadynamics \cite{martovnak2005simulation}, quasirandom structure search \cite{pickard2011ab} and evolutionary algorithm \cite{glass2006uspex}. Among all, evolutionary algorithms are widely used in structure prediction since they are population-based, can find various global and local minima with various initial guesses, and often show more robust searching without being trapped in local minima. 

Existing evolutionary strategies generally involve two key steps: the initialization of structure population for given chemical composition and the update of the population after the evaluation of target property. The evolutionary algorithm is iterative and for the start it requires creation of an initial population of structures. It is realized by special quasirandom initialization \cite{lyakhov2013new}, which does not rely on any knowledge about already generated structures. 

The use of previously generated structures is significant for the task of cluster structure prediction, where the variable, responsible for the number of atoms in a cluster, is added to the stoichiometric variable. It appears feasible that clusters with close number of atoms belong to the same structural motif \cite{FlikkemaBromley2004, BromleyStefan2005, ignatov2021many}. Therefore a method allowing to predict structures with use of already known structures is of interest.

Deep learning methods are capable of predicting accurate properties and potential energy of chemical compounds \cite{schutt2017schnet, klicpera_dimenet_2020, klicpera2021gemnet}. In particular, generative machine learning models are a powerful data-driven approach to sampling from a learned distribution of molecular configuration. Furthermore, some generative models allow to sample 3d molecular configurations, that is one of the most informative way of representation. Models that enable sampling of 3d configurations can be divided into autoregressive \cite{gebauer2018generating}, \cite{gebauer2020symmetryadapted}, \cite{LiStructure2021}, \cite{gebauer2021inverse}, generative adversarial network \cite{hoffmann2019generating}, variational autoencoder \cite{nesterov20203dmolnet}, reinforcement learning \cite{simm2020reinforcement}, \cite{simm2020symmetryaware}, \cite{meldgaard2021generating}, equivariant normalizing flows \cite{satorras2021en}, equivariant diffusion models \cite{hoogeboom2022equivariant}. 

Molecules are stable compounds whose atoms are covalently bonded to each other. In contrast, clusters are a loosely organized group of atoms that may have a much more complex (and diverse) structure. In result, it is complicated to apply the majority of molecules generation approaches to cluster generation. It seems the only approaches are suitable that generate the whole structure at the same time. The energy-based models are one of straightforward way to approach generation. They allow to access diverse samples, admit a naive structure representation, and the most important - permit physical interpretation of energy function that may also be called an artificial interatomic potential surface. This specific artificial potential energy can be interpreted as 'smoothed' version of the actual interatomic potential surface. The both of these surfaces have sustainable low-energy cluster structures as local minima with an equal energy. On the other hand, artificial potential surface has no local minima, corresponding with high-energy cluster structures. This option significantly simplifies cluster generation. 

In this work we establish an energy-based model and generate silica clusters. An artificial potential of the output satisfies two requirements: the value of the potential coincides with the physical value of the energy for stable structures, whereas the stable structure is a local minimum of this potential surface. The application of the indicated approach to silica clusters $(SiO_2)_n$ has resulted in the discovery of the stable structures with $n = 28\ldots 51$.

\section{Methods}

The method is to built a model that describes a potential surface where each cluster structure, used in training the model, is a local minimum on this surface. To train the model, clusters with different number of atoms are taken. For each chemical compound of the cluster, structures are selected with energies that do not differ significantly from the energy of the global minimum. Unlike the potential surface described using DFT, quantum empirical methods, or their approximations \cite{tsuneyuki1988first, van1990force, flikkema2003new, FlikkemaBromley2009}, the surface constructed through the proposed method has a smaller number of local minima. This property of the potential surface is also hold for clusters with more atoms than were used in training. 

The dominance of low-energy local minima on the potential surface for clusters with a large number of atoms can be explained by the idea that the model learns the specific patterns for low-energy structures and extrapolates them. Based on that, it additionally follows that the method cannot generate structures that do not have common patterns with structures with fewer atoms. Structure generation consists in finding local minima of the potential surface, for which the Langevin \cite{GM91} algorithm is used. 

The description of the method is divided into the description of the potential model training, section~\ref{sec:artificial_potential_surface}, and the description of generation process using the trained potential,  section~\ref{sec:generation}.

\subsection{Artificial potential surface}
\label{sec:artificial_potential_surface}

The training data consists solely of stable cluster structures and their energies. Using this data, it is possible to determine the values of the potential surface $U$ for the equilibrium structure of cluster. To describe the potential surface for non-equilibrium structures of cluster atoms, the fact of stability of the initial structure is used: the forces, acting on randomly displaced atoms of the cluster, point in the direction of the equilibrium position of the atoms.

Descriptions of the energy surface and forces acting on atoms are possible with a variety of models \cite{schutt2017schnet, klicpera_dimenet_2020, klicpera2021gemnet}. The chosen architecture of SchNet \cite{schutt2017schnet} is relatively simple but powerful, in addition further changes have been made to the architecture. The main differences from the original SchNet: layer norm \cite{ba2016layer} for learning stability, variational dropout \cite{NeklyudovBayesianPruning} to prevent overfitting, the activation function LeakySiLU (\ref{leakySiLU}) - a differentiable analogue of LeakyReLU, radius base functions are implemented by trigonometric functions. For a more detailed description of the architecture used, see Appendix~\ref{appendix:model}.

The cluster structure is described by determining the atomic charges $Z=(z_1, \ldots, z_n)$ and their coordinates $R=(r_1, \ldots,r_n)$, where $n$ is the number of cluster atoms. The set of these quantities is denoted by $X=(Z, R)$. When an atom is taken out of its equilibrium position, it changes only the matrix of atomic coordinates. The charge matrix is always constant. Denote by $X_0$ - an arbitrary stable structure from the training sample, $U_0$ - the energy corresponding to it. In the learning process, for each structure, its deformation $X_\sigma\sim\mathcal{N}(X|X_0,\sigma^2\mathbb{I})$ is randomly constructed, where the value of deformation $\sigma$ is distributed according to the law $ p(\sigma)\propto 1/\sigma$, $\sigma\in\left[\sigma_{\min}, \sigma_{\max}\right]$. The range of taken $\sigma$ values is based on the notion that $\sigma_{\max}\sim L$, where $L=1\text{Å}$ is the characteristic interatomic distance, and $\sigma_{\min} \ll L$.

Finding the weights of the artificial potential model is carried out by minimizing the loss function
\begin{equation}\label{model_loss}
    \mathcal{L} = w_{property}\mathcal{L}_{property} + w_{cosine}\mathcal{L}_{cosine} + w_2\mathcal{L}_2
\end{equation} 
which is the weighted sum of the term describing the energy
\begin{equation}\label{loss:property}
\mathcal{L}_{property} = \mathbb{E}_{X_0}\|U(X_0) - U_0\|^2
\end{equation}
term describing the direction of the gradients of the potential
\begin{equation}\label{loss:cosine}
\mathcal{L}_{cosine} = \mathbb{E}_{X_0}\mathbb{E}_{\sigma\sim p(\sigma)}\mathbb{E}_{X_\sigma|X_0}\left[1 - \frac{\nabla U(X_\sigma)\cdot(X_\sigma - X_0)}{\|\nabla U(X_\sigma)\| \|X_\sigma - X_0\|}\right]
\end{equation}
and a term describing the stability of the model in the equilibrium region
\begin{equation}\label{loss:l2}
    \mathcal{L}_{2} = \|\nabla U(X_\sigma) - (X_\sigma - X_0)\|^2.
\end{equation}
The weights $w_{property}$ and $w_{cosine}$ are chosen approximately equal, and the value $w_2\ll \min(w_{property}, w_{cosine})$. The loss function (\ref{loss:l2}) is used solely to eliminate the instability of potential energy gradients in the equilibrium region \cite{du2020implicit}.

In general, the graph convolution networks are not able to generalize from small to large graphs \cite{yehudai2021local}. In the case of the cluster generation, the more size of the structure, the less often the model finds a good local minima. This feature can be indicated with a concept of a threshold number of atoms in a cluster, above which the model does not find a good local minima. The threshold value depends on the size of the clusters used to build the model and can be increased by use of the bulk-cut nanocrystal \cite{persson2003smallest}.

\subsection{Generation}
\label{sec:generation}

Structure generation implies finding a local minimum of the potential surface $U(x)$. Finding a local minimum is carried out by the Langevin algorithm \cite{GM91}
\begin{equation}\label{algorithm:langeving:equation}
\begin{split}
    &R_{k+1} = R_k - \alpha_k\nabla_R \widetilde{U}(X_k) + \beta_k\xi,\\ 
    &Z_{k+1} = Z_k \\
    &\xi\sim\mathcal{N}(0, \mathbb{I}),\quad k=0,\ldots,N_g-1
\end{split}
\end{equation}
here $N_g$ is a number of generation steps, $\widetilde{U}(X)$ is a potential function that is the sum of the artificial potential $U(x)$ and the regularization potential $\delta U$ described below. The initial state $X_0=(Z_0, R_0)$ consists of a given charge vector $Z_0$ and the positions of the atoms sampled from normal distribution $R_0\sim\mathcal{N}(0, \Sigma^2_n\mathbb{I})$, where $\Sigma^2_n$ - the variance of the position of atoms depending on the number of atoms $n$ is determined by the training set. Exponential parameter decay rate is used 
\begin{equation}\label{algorithm:langeving:parameters}
    \alpha_k = \alpha_{\max}\left(\frac{\alpha_{\min}}{\alpha_{\max}}\right)^{k/(N_g-1)},\quad
    \beta_k = \beta_{\max}\left(\frac{\beta_{\min}}{\beta_{\max}}\right)^{k/(N_g-1)}
\end{equation}
Good to note that the exponential decay rate does not guarantee finding the global minimum, but according to our observations it leads to the generation of rather good structures in a small number of steps. We associate the noise parameters with the characteristic parameters of the interatomic distance, namely, $\beta_{\max}\sim 1\text{Å}$ and $\beta_{\min}\ll 1\text{Å}$. The ratios $\alpha_{\min}$ and $\alpha_{\max}$ depend on the potential and require a selection for each potential from the condition for obtaining qualitative structures.

One of the conditions for the Langevin algorithm to find the global minimum of the potential $U(X)$ is its asymptotic behavior \cite{GM91}
\begin{equation}\label{potential:assymptotic}
    U(X)\rightarrow\infty \ \text{and}\ |\nabla_R U(X)|\rightarrow\infty\quad \text{if} \ |R|\rightarrow\infty
\end{equation}
This condition is violated by any deep learning model. Any model is defined on the structures of clusters whose atoms rest in some finite volume. This limitation arises due to the fact that the training uses clusters that do not describe all possible volumes. Besides that, it arises when we use the cutoff thresholds by distance in model architectures. Violation of the condition (\ref{potential:assymptotic}) is observed as the generation of an unstructured cloud of atoms or separate groups of structures.

This problem is solved by introducing a regularizing potential $\delta U(x)$ satisfying the property (\ref{potential:assymptotic}). Furthermore, the potential $\delta U(X)$ should slightly distort $U(X)$ on the cluster structures and eventually can be neglected when training the artificial potential.
The algorithm below is designed for constructing the regularizing potential $\delta U(x)$. The symbol $\mathcal{N}$ denotes the set of different atoms that make up the structure. The structure in our algorithm is represented as a fully connected weighted graph $\mathcal{G}$. The nodes of the graph are the atoms of the structure. The edge connecting the atom $i$ with the atom $j$ has the weight $r_{ij}$ corresponding to the distance between these atoms. For each pair of atoms $I, J\in\mathcal{N}$, found in the graph $\mathcal{G}$, we leave only the edges connecting these atoms. On the resulting subgraph, we construct a minimal spanning tree and denote the edges of this tree by $\mathcal{E}_{IJ}$. If the weight distribution of $\mathcal{E}_{IJ}$ edges is described by a unimodular distribution, the potential energy is defined 
\begin{equation}\label{reg:mst_ij}
\begin{split}
    \delta U_{IJ}(X)&=\sum\limits_{i,j\in \mathcal{E}_{IJ}}\left(2\log (r_{ij}+\epsilon)-\frac{(r_{ij}-\zeta_{IJ})^2}{2\eta^2_{IJ}}\right),\\ \zeta_{IJ} &= \frac{R^2_{IJ}-2\eta^2_{IJ}}{R_{IJ}}, \ \epsilon \ll R_{IJ}
\end{split}    
\end{equation}
where $R_{IJ}$ is the average weight of the edge $\mathcal{E}_{IJ}$, $\eta_{IJ}$ is the characteristic standard deviation of the weights of the edges $\mathcal{E}_{IJ}$. The regularizing potential $\delta U(x)$ for the structure is defined as the total of the potentials for each pair of atoms $I$ and $J$. 
\begin{equation}\label{reg:mst}
    \delta U(X)=\frac{C}{2}\sum\limits_{I,J\in \mathcal{N}}\delta U_{IJ}(X)
\end{equation}
here $C$ is a constant. The list of pairs of atoms used in the potential (\ref{reg:mst}) and the value of $R_{IJ}$ are determined on the cluster structures used for learning. The values of $C$ and $\eta_{IJ}$ are identified from the analysis of the structures obtained during generation.

It should be noted that the potential (\ref{reg:mst}) does not provide the conditions for finding the global minimum \cite{GM91}, which manifests itself in the appearance of linear structures during generation. The $\log r$ term in the function (\ref{reg:mst_ij}) can make the Langevin algorithm unstable. The advantage of the potential (\ref{reg:mst}) is that it practically disappears for stable structures, since it does not require selection of parameters depending on the number of atoms.

\section{Experiments}
\label{experiments}

The method is examined on the example of the generation of silicon oxide cluster structures. Training examples with composition $Si_{n}O_{2n\pm1}$, $n=10,\ldots , 27$ are obtained by cluster optimization in USPEX code \cite{oganov2006crystal, lyakhov2013new, lepeshkin2018method}. In addition, we use the structures that have been published in the literature \cite{FlikkemaBromley2004, BromleyStefan2005, FlikkemaBromley2009}. Two models of artificial potential are trained. The first model is trained on cluster structures with the chemical composition $Si_{n}O_{2n\pm1}$, $n=10,\ldots , 21$. The search for structure clusters is carried out for the compositions $Si_{n}O_{2n}$, $n=22,\ldots , 30$. The generated structures are compared with those found by USPEX and published in the literature. The second model is trained on cluster structures with the chemical composition $Si_{n}O_{2n\pm1}$, $n=10,\ldots , 30$. The search for cluster structures is carried out for chemical compositions $Si_{n}O_{2n}$, $n=28,\ldots , 50$. As a result, new structures of silicon oxide clusters are proposed. Clusters of this size have not been previously considered in the literature.

\subsection{DFT approximation with MNDO}
\label{subsection:mndo}

\begin{figure}
    \centering
    \includegraphics[totalheight=0.5\textwidth]{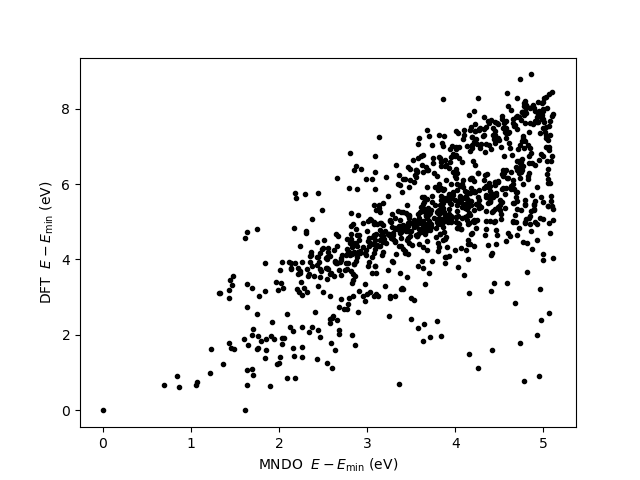}
    \caption{Compare MNDO/MOPAC with DFT/Gaussian energy calculation for $Si_{15}O_{30}$ isomers.}
    \label{fig:dft_vs_mndo}
\end{figure}

During our previous calculations \cite{nanoscale2016, lepeshkin2018method} we found that $Si-O$ clusters are well described within the semi-empirical MNDO approach that is implemented, for example, in the MOPAC package \cite{dewar1985development}. This is confirmed by figure~\ref{fig:dft_vs_mndo} where the energy of the first 1000 isomers, found during the global optimization of $Si_{15}O_{30}$ cluster within MNDO approach, is demonstrated along the energies of the corresponding structures, recalculated within the DFT approach (using B3LYP/6-311+G(d,p) implemented in Gaussian code \cite{stephens1994ab, frisch2009gaussian}). Therefore, applying this approach can significantly reduce the relaxation time of one structure and allows to generate a huge number of “realistic” (low-energy) structures of silicon-oxide clusters.

\subsection{Training samples}
\label{subsection:train_dataset}

\begin{figure}[H]
    \centering
    \begin{subfigure}{0.25\textwidth}
        \centering
        \includegraphics[width=\textwidth]{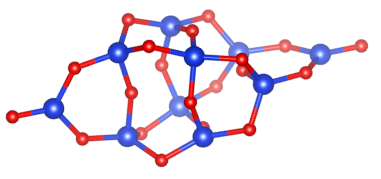}
        \caption*{$Si_{10} O_{20}$}
    \end{subfigure}
    \begin{subfigure}{0.25\textwidth}
        \centering
        \includegraphics[width=\textwidth]{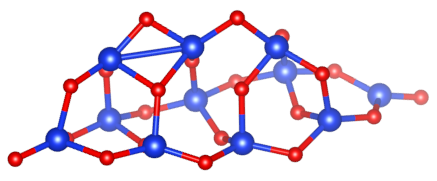}
        \caption*{$Si_{11} O_{22}$}
    \end{subfigure}
    \begin{subfigure}{0.25\textwidth}
        \centering
        \includegraphics[width=\textwidth]{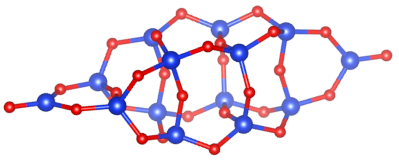}
        \caption*{$Si_{14} O_{28}$}
    \end{subfigure}
    \begin{subfigure}{0.25\textwidth}
        \centering
        \includegraphics[width=\textwidth]{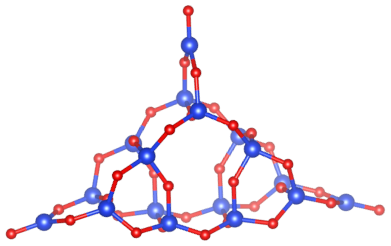}
        \caption*{$Si_{17} O_{34}$}
    \end{subfigure}
    \begin{subfigure}{0.25\textwidth}
        \centering
        \includegraphics[width=\textwidth]{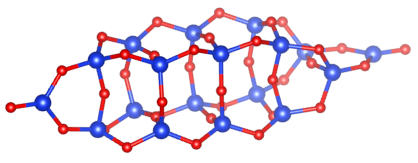}
        \caption*{$Si_{18} O_{36}$}
    \end{subfigure}
    \begin{subfigure}{0.25\textwidth}
        \centering
        \includegraphics[width=\textwidth]{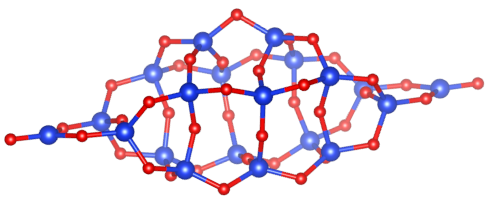}
        \caption*{$Si_{20} O_{40}$}
    \end{subfigure}
    \begin{subfigure}{0.26\textwidth}
        \centering
        \includegraphics[width=\textwidth]{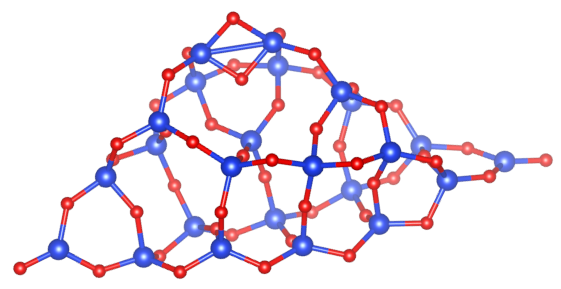}
        \caption*{$Si_{24} O_{48}$}
    \end{subfigure}
    \begin{subfigure}{0.20\textwidth}
        \centering
        \includegraphics[width=\textwidth]{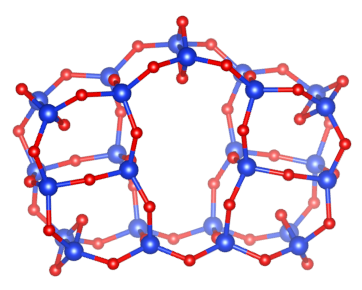}
        \caption*{$Si_{26} O_{52}$}
    \end{subfigure}
    \begin{subfigure}{0.22\textwidth}
        \centering
        \includegraphics[width=\textwidth]{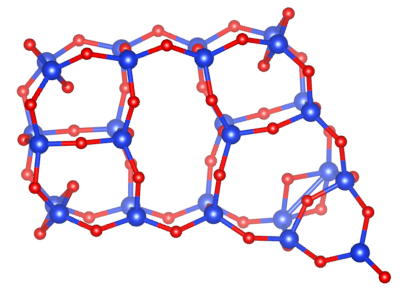}
        \caption*{$Si_{27} O_{54}$}
    \end{subfigure}
    \caption{Common motifs of structures used to train the artificial potential model}
    \label{fig:train}
\end{figure}

The training examples are composed of cluster structures with chemical compositions $Si_{n}O_{2n\pm1}$, $n=10,\ldots , 27$ generated by the variable-composition evolutionary algorithm implemented in the USPEX code \cite{oganov2006crystal, lyakhov2013new, lepeshkin2018method}, combined with the semi-empirical MNDO approach as implemented in the MOPAC package. For each  composition, cluster structures were taken with energies not exceeding the minimum energy of more than 6 eV. The total number of structures in this sample reach 22500. Figure~\ref{fig:train} shows typical structures obtained. Bulk-cut nanocrystal with the chemical composition $Si_{95}O_{205}$ was used to regularize the model. It was obtained from a fragment of a silicon oxide crystal by cutting out a ball.

\subsection{Parameters of the artificial potential model}

The parameters of the model are chosen to be $N=6$ interacting blocks, $H=4$ heads in each block, $D=256$ dimension of the hidden vectors of representation of atoms, $FF=256$ dimension of the inner Atom-wise feedforward layer. Distance encoding is carried out using $K=32$ $\sin$ functions and the same number of $\cos$ functions, the scale parameter is chosen equal to $R_{rbf}=15.0 \text{Å}$, the frequency range from $f_{\ min}=0.125$ to $f_{\max}=8.0$. We take $R_{cut} = 7.5\text{Å}$ as the distance cutoff threshold with a slope of $S_{cut}=2.0$. To train the model, RMSProp optimization algorithm is used.

The training of the artificial potential model is divided into several steps. Firstly, the model is trained to describe only the energy of clusters using the loss function (\ref{loss:property}). Secondly, the task of restoring the cluster structure, the loss function (\ref{loss:cosine}), is added. Finally, jumps in the modulus of the gradient around the local minimum, the loss function (\ref{loss:l2}), are removed.

\subsection{Generation of cluster structures with composition from $Si_{22}O_{44}$ to $Si_{30}O_{60}$}
\label{subsection:Si_22_Si_30}

The training examples are composed only of the structures described in section~\ref{subsection:train_dataset} with chemical composition $Si_{n}O_{2n\pm1}$, where $n=10, \ldots, 21$. The energies of the selected clusters do not exceed the minimum energy of more than 4 eV. The total number of structures used for training clusters is $15\cdot 10^3$.

Approximately $74\cdot 10^3$ generations of cluster structures are carried out for the chemical compositions $Si_{n}O_{2n}$, $n=22,\ldots , 30$. All structures found using the artificial potential are optimized using MNDO/MOPAC. For each composition, the 100 lowest energy structures are selected and these structures are optimized by USPEX code for 5 to 10 generations to eliminate defects.

\begin{figure}[ht]
    \centering
    \begin{subfigure}{0.3\textwidth}
        \centering
        \includegraphics[width=\textwidth]{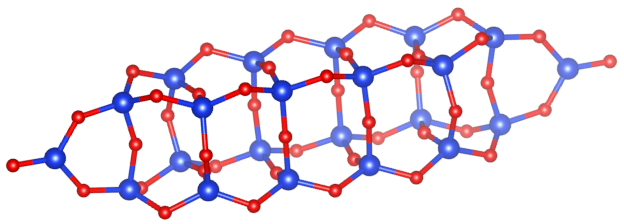}
        \caption*{$Si_{22} O_{44}$}
    \end{subfigure}
    \begin{subfigure}{0.3\textwidth}
        \centering
        \includegraphics[width=\textwidth]{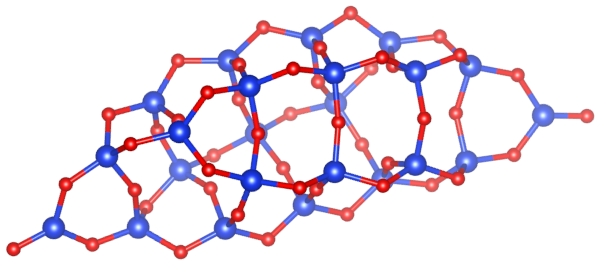}
        \caption*{$Si_{23} O_{46}$}
    \end{subfigure}
    \begin{subfigure}{0.3\textwidth}
        \centering
        \includegraphics[width=\textwidth]{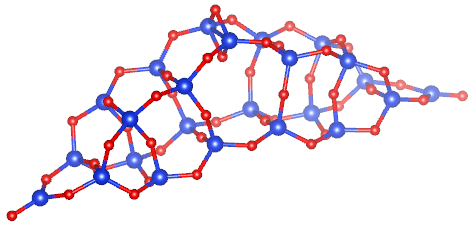}
        \caption*{$Si_{24} O_{48}$}
    \end{subfigure}
    \centering
    \begin{subfigure}{0.3\textwidth}
        \centering
        \includegraphics[width=\textwidth]{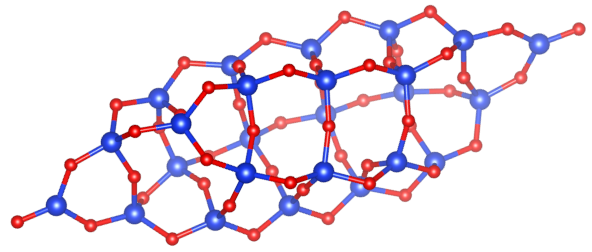}
        \caption*{$Si_{25} O_{50}$}
    \end{subfigure}
    \centering
    \begin{subfigure}{0.3\textwidth}
        \centering
        \includegraphics[width=\textwidth]{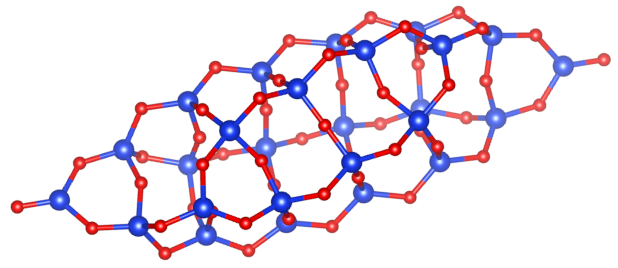}
        \caption*{$Si_{26} O_{52}$}
    \end{subfigure}
    \begin{subfigure}{0.3\textwidth}
        \centering
        \includegraphics[width=\textwidth]{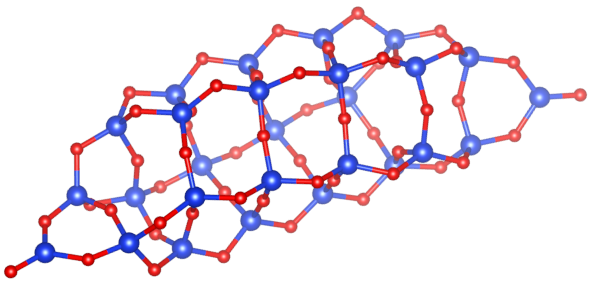}
        \caption*{$Si_{27} O_{54}$}
    \end{subfigure}
    \centering
    \begin{subfigure}{0.3\textwidth}
        \centering
        \includegraphics[width=\textwidth]{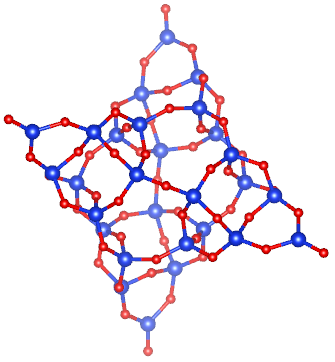}
        \caption*{$Si_{28} O_{56}$}
    \end{subfigure}
    \begin{subfigure}{0.3\textwidth}
        \centering
        \includegraphics[width=\textwidth]{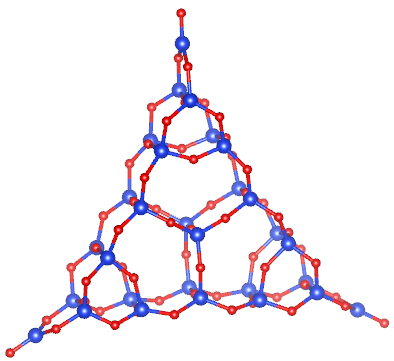}
        \caption*{$Si_{29} O_{58}$}
    \end{subfigure}
    \begin{subfigure}{0.3\textwidth}
        \centering
        \includegraphics[width=\textwidth]{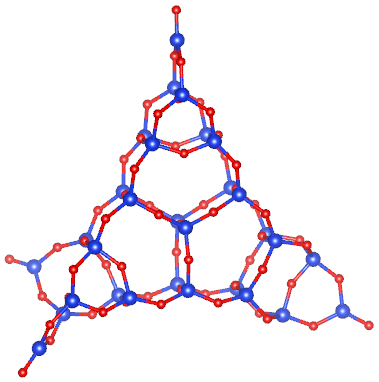}
        \caption*{$Si_{30} O_{60}$}
    \end{subfigure}
    \caption{Examples of clusters found by the model trained on $Si_{n}O_{2n\pm1}$, $n=10, \ldots, 21$}
    \label{fig:gi2}    
\end{figure}

Examples of the retrieved cluster structures are shown in figure~\ref{fig:gi2}. It appears that these structures are an extrapolation of the examples used to train the potential (figure~\ref{fig:train}). All found structures are less energetically favorable than the structures given in \cite{BromleyStefan2005}, except for $Si_{22}O_{44}$. Table~\ref{tab:energy_cmp} confronts the energies of the generated structures with those found during the USPEX code optimization (section~\ref{subsection:train_dataset}). The comparison indicates that the artificial potential has found structures that are more advantageous in terms of energy.

In the paper \cite{BromleyStefan2005} shows that up to $Si_{22}O_{44}$ the global minima is formed by columnar-like structures. Starting from $Si_{23}O_{46}$ to $Si_{27}O_{54}$, disk-like structures form the global minimum. The training sample was composed of structures up to $Si_{21}O_{42}$ (figure~\ref{fig:train}), hence, from columnar-like structures. The method extrapolated this form up to $Si_{27}O_{54}$. Thus, one global minimum $Si_{22}O_{44}$, already proposed in \cite{BromleyStefan2005}, was indicated. However, disk-like structures were not found by the method. The exception is $Si_{29}O_{56}$ (figure~\ref{fig:gi2}) that could be extrapolated from $Si_{17}O_{34}$ (figure~\ref{fig:train} ). The fact that there are no disk-like structures demonstrates the limitations of the method. Therefore if characteristic patterns are absent in the training set, they will not be found for structures with a large number of atoms.

The advantage of the method is its versatility. It does not require the development of special potentials for each chemical composition of the cluster \cite{tsuneyuki1988first, van1990force, flikkema2003new, FlikkemaBromley2009}. The fact that the method extrapolates patterns that have appeared to be optimal on clusters with fewer atoms allows to propose more advantageous structures in less computational time. This is confirmed by the fact that the method found more optimal structures than USPEX code.

\begin{table}[h!]
    \centering
    \begin{tabular}{|c|c|c|}
    \hline
    Formula & USPEX code (eV) & Model (eV) \\
    \hline
    $Si_{22}O_{44}$ & -16198.153 & -16202.7047 \\
    \hline
    $Si_{23}O_{46}$ & -16935.2157 & -16938.18108 \\
    \hline
    $Si_{24}O_{48}$ & -17674.42 & -17675.00966 \\
    \hline
    $Si_{25}O_{50}$ & -18411.1303 & -18412.54393 \\
    \hline
    $Si_{26}O_{52}$ & -19149.0923 & -19149.72417 \\
    \hline
    $Si_{27}O_{54}$ & -19886.1902 & -19886.9401 \\
    \hline    
    \end{tabular}
    \caption{Comparison of structures energies found by the USPEX code and the model. The energies are calculated using the semi-empirical MNDO method implemented in the MOPAC package.}
    \label{tab:energy_cmp}
\end{table}

\subsection{Generation of cluster structures with composition from $Si_{28}O_{56}$ to $Si_{50}O_{100}$}

The training examples are composed of the cluster structures described in section~\ref{subsection:train_dataset} with chemical formulas $Si_{n}O_{2n\pm1}$, $n=10,\ldots , 27$.
992 structures with chemical composition $Si_{n}O_{2n\pm1}$, $n=22,\ldots , 30$ obtained in the section~\ref{subsection:Si_22_Si_30} are added to these training examples. For each atomic composition, cluster structures are taken with energies not exceeding the minimum energy of more than 3 eV.
In addition, clusters with chemical formulas $Si_nO_{2n}$ where $n=10, \ldots, 27$ described in \cite{FlikkemaBromley2004, BromleyStefan2005, FlikkemaBromley2009} papers, which were obtained using the heuristic potential \cite{flikkema2003new, FlikkemaBromley2009} and the basing hopping \cite{basinhopping1997} global minimum search algorithm. These clusters are global minima for each composition. The cluster energies are calculated using MNDO. The total number of structures in this sample is 19.

\begin{figure}[H]
    \centering

    \begin{subfigure}{0.2\textwidth}
        \centering
        \includegraphics[width=\textwidth]{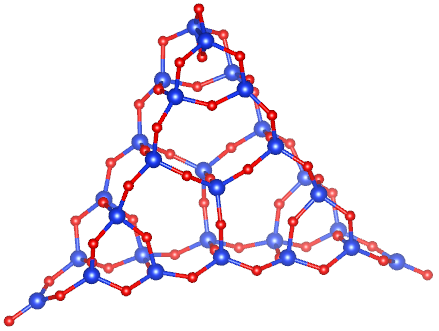}
        \caption*{$Si_{28} O_{56}$}
    \end{subfigure}
    \begin{subfigure}{0.2\textwidth}
        \centering
        \includegraphics[width=\textwidth]{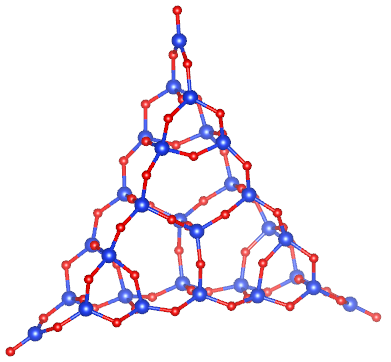}
        \caption*{$Si_{29} O_{58}$}
    \end{subfigure}
    \begin{subfigure}{0.2\textwidth}
        \centering
        \includegraphics[width=\textwidth]{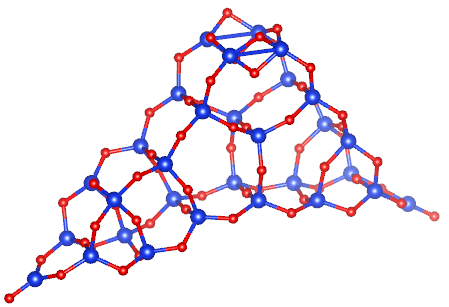}
        \caption*{$Si_{30} O_{60}$}
    \end{subfigure}
    \begin{subfigure}{0.2\textwidth}
        \centering
        \includegraphics[width=\textwidth]{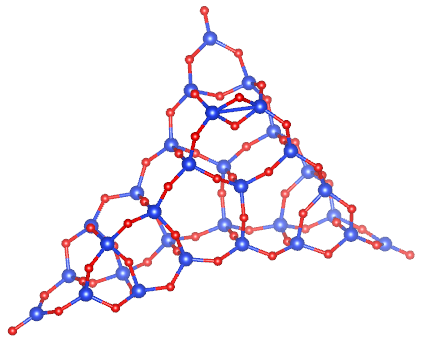}
        \caption*{$Si_{31} O_{62}$}
    \end{subfigure}
    
    \begin{subfigure}{0.2\textwidth}
        \centering
        \includegraphics[width=\textwidth]{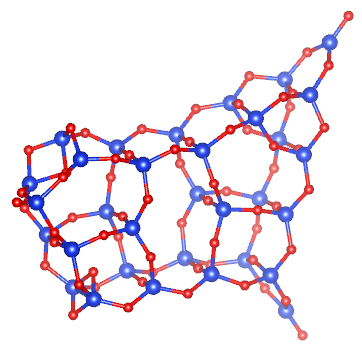}
        \caption*{$Si_{32} O_{64}$}
    \end{subfigure}
    \begin{subfigure}{0.2\textwidth}
        \centering
        \includegraphics[width=\textwidth]{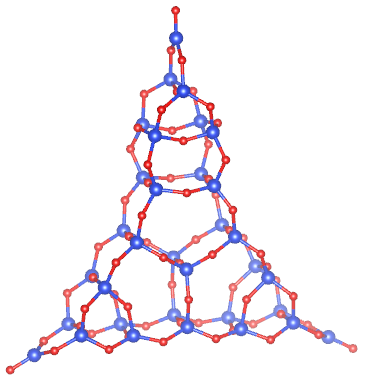}
        \caption*{$Si_{33} O_{66}$}
    \end{subfigure}
    \begin{subfigure}{0.2\textwidth}
        \centering
        \includegraphics[width=\textwidth]{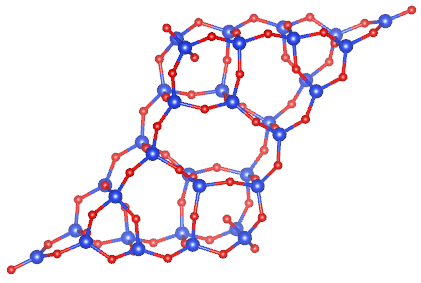}
        \caption*{$Si_{34} O_{68}$}
    \end{subfigure}
    \begin{subfigure}{0.2\textwidth}
        \centering
        \includegraphics[width=\textwidth]{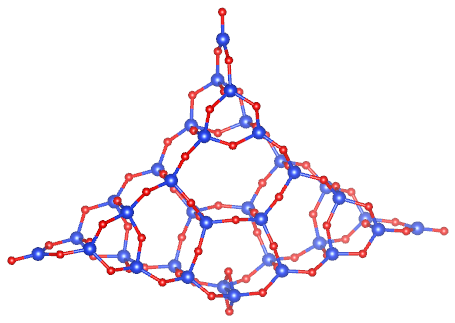}
        \caption*{$Si_{35} O_{70}$}
    \end{subfigure}

    \begin{subfigure}{0.2\textwidth}
        \centering
        \includegraphics[width=\textwidth]{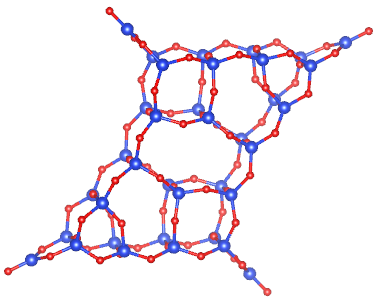}
        \caption*{$Si_{36} O_{72}$}
    \end{subfigure}
    \begin{subfigure}{0.2\textwidth}
        \centering
        \includegraphics[width=\textwidth]{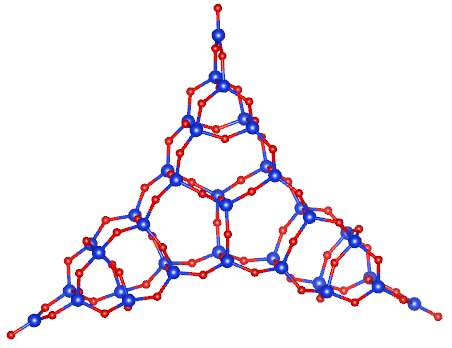}
        \caption*{$Si_{37} O_{74}$}
    \end{subfigure}
    \begin{subfigure}{0.2\textwidth}
        \centering
        \includegraphics[width=\textwidth]{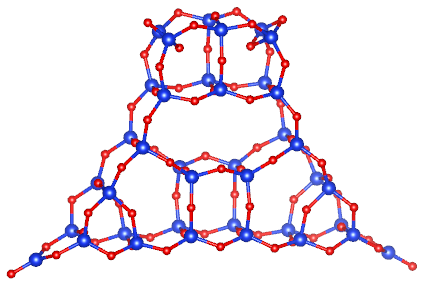}
        \caption*{$Si_{38} O_{76}$}
    \end{subfigure}
    \begin{subfigure}{0.2\textwidth}
        \centering
        \includegraphics[width=\textwidth]{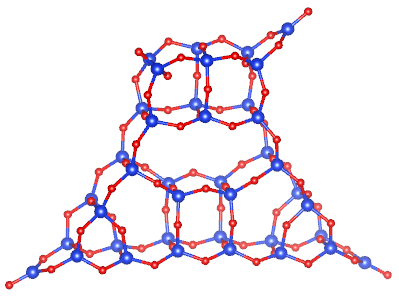}
        \caption*{$Si_{39} O_{78}$}
    \end{subfigure}

    \begin{subfigure}{0.25\textwidth}
        \centering
        \includegraphics[width=\textwidth]{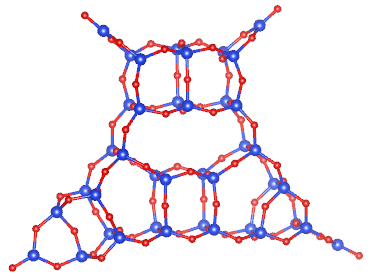}
        \caption*{$Si_{40} O_{80}$}
    \end{subfigure}
    \begin{subfigure}{0.15\textwidth}
        \centering
        \includegraphics[width=\textwidth]{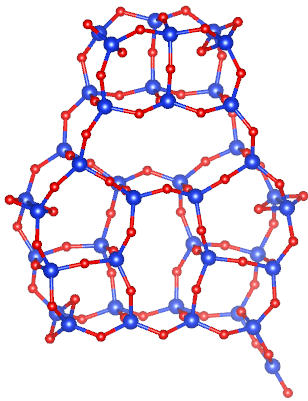}
        \caption*{$Si_{41} O_{82}$}
    \end{subfigure}
    \begin{subfigure}{0.15\textwidth}
        \centering
        \includegraphics[width=\textwidth]{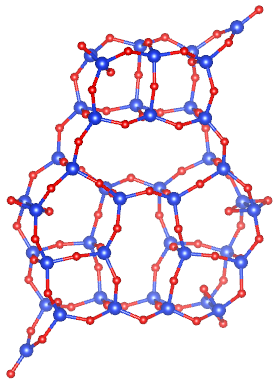}
        \caption*{$Si_{42} O_{84}$}
    \end{subfigure}
    \begin{subfigure}{0.15\textwidth}
        \centering
        \includegraphics[width=\textwidth]{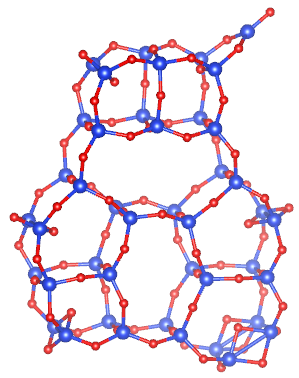}
        \caption*{$Si_{43} O_{86}$}
    \end{subfigure}

    \begin{subfigure}{0.17\textwidth}
        \centering
        \includegraphics[width=\textwidth]{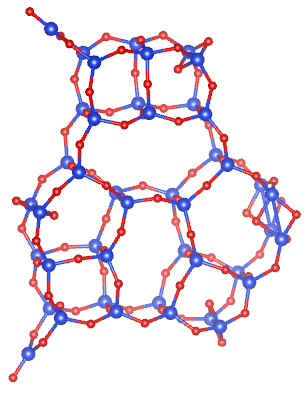}
        \caption*{$Si_{44} O_{88}$}
    \end{subfigure}
    \begin{subfigure}{0.17\textwidth}
        \centering
        \includegraphics[width=\textwidth]{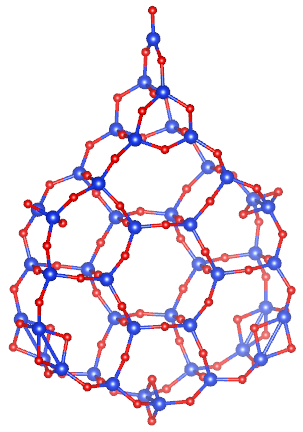}
        \caption*{$Si_{45} O_{90}$}
    \end{subfigure}
    \begin{subfigure}{0.17\textwidth}
        \centering
        \includegraphics[width=\textwidth]{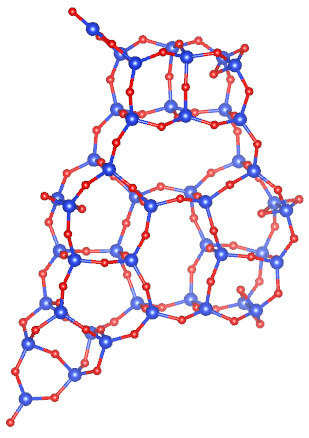}
        \caption*{$Si_{46} O_{92}$}
    \end{subfigure}
    \begin{subfigure}{0.17\textwidth}
        \centering
        \includegraphics[width=\textwidth]{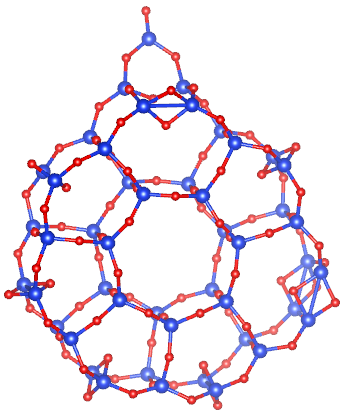}
        \caption*{$Si_{47} O_{94}$}
    \end{subfigure}

    \begin{subfigure}{0.17\textwidth}
        \centering
        \includegraphics[width=\textwidth]{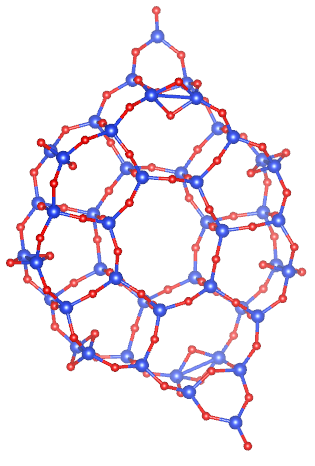}
        \caption*{$Si_{48} O_{96}$}
    \end{subfigure}
    \begin{subfigure}{0.17\textwidth}
        \centering
        \includegraphics[width=\textwidth]{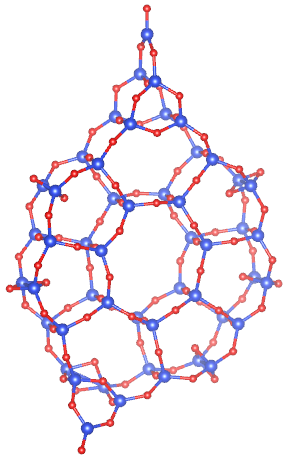}
        \caption*{$Si_{50} O_{100}$}
    \end{subfigure}
    \begin{subfigure}{0.25\textwidth}
        \centering
        \includegraphics[width=\textwidth]{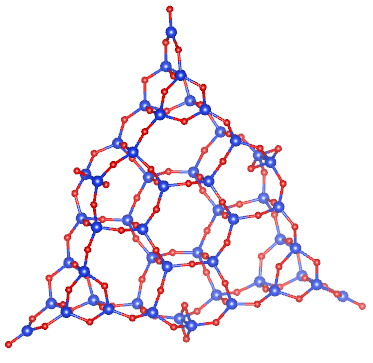}
        \caption*{$Si_{51} O_{102}$}
    \end{subfigure}
    \caption{Examples of clusters found by the model trained on $Si_{n}O_{2n\pm1}$, $n=10, \ldots, 30$}
    \label{fig:gi34}
\end{figure}

Approximately $120\cdot 10^3$ generations of cluster structures are carried out for the chemical compositions $Si_{n}O_{2n}$, $n=28,\ldots , 50$. All structures found using the artificial potential are optimized using MNDO. For a number of structures, the geometry is changed manually. For each chemical composition, several structures with the lowest energy are selected. These structures are optimized by Gaussian B3LYP/6-31G(d,p) \cite{stephens1994ab, frisch2009gaussian}. Figure~\ref{fig:gi34} shows structures with the lowest energy.

From the examples shown in figure~\ref{fig:gi34}, we can conclude that the artificial potential model allows us to remember and extrapolate the patterns characteristic of low-energy structures. In particular, the obtained $Si_{28}O_{56}$, $\ldots$, $Si_{31}O_{62}$ can be interpreted as an extrapolation of the structures found in \cite{BromleyStefan2005}. The $Si_{36}O_{72}$ and $Si_{45}O_{90}$ structures, for example, represent an extrapolation of the $Si_{27}O_{54}$ structure found by USPEX code optimization.

\section{Conclusion}

We apply energy-base model in combination with graph convolutional networks to construct artificial interatomic potential that reproduces ground-state and low-energy structures correctly. It is significantly smoother than "real" potential in the intermediate regions of the configuration space. Carrying out global optimization in such efficient energy landscape gives much faster convergence compared to the correct (DFT) landscape. Implementation of this approach requires it the initial generation of a sufficient number of good cluster structures, for which other methods of structure generation should be used (e.g. evolutionary algorithms). The model allows to generate larger clusters than ones in the training set.

We applied this approach to silica clusters $(SiO_2)_n$. The model was trained on structures calculated by us and from literature data containing up to n=27 silicon atoms. As a result, we propose novel ground-state candidate structures of $(SiO_2)_n$ clusters with $n$ up to 51.

\printbibliography

\appendix
\section*{Appendix}

\section{Model architecture}
\label{appendix:model}

\begin{figure}[H]
    \centering
    \begin{subfigure}{0.33\textwidth}
        \centering
        \includegraphics[width=\textwidth]{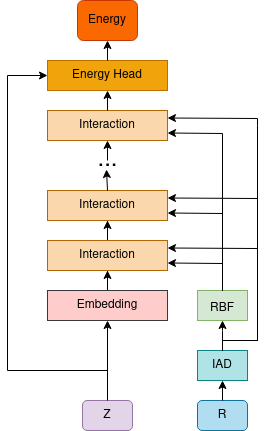}
        \caption{}
        \label{fig:total_model_scheme}
    \end{subfigure}
    \begin{subfigure}{0.28\textwidth}
        \centering
        \includegraphics[width=\textwidth]{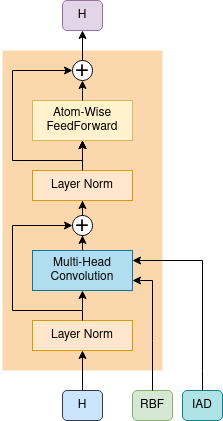}
        \caption{}
        \label{fig:interaction_model_scheme}
    \end{subfigure}
    \begin{subfigure}{0.27\textwidth}
        \centering
        \includegraphics[width=\textwidth]{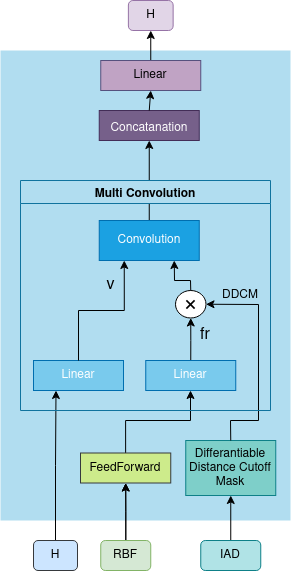}
        \caption{}
        \label{fig:convolution_model_scheme}
    \end{subfigure}
    \caption{Model architecture}
\end{figure}

The architecture of the artificial potential model is similar to the architecture of SchNet \cite{schutt2017schnet}. The principal design of the model is shown in figure~\ref{fig:total_model_scheme}, figure~\ref{fig:interaction_model_scheme} shows the interaction block diagram, figure~\ref{fig:convolution_model_scheme} shows the convolution block diagram.

Each structure is represented as a set of $n$ atoms with their charges $Z=(z_1, \ldots, z_n)$ and their positions $R=(r_1,\ldots, r_n)$. Each atom in the model is represented as a vector $h^l_i\in\mathbb{R}^D$, where $l=1, \ldots, N$ is the layer number, $i=1\ldots n$ is the atom number, $N$ is the number of interaction layers, $D$ is the dimension of the presentation space. The structure is described as a matrix composed of representation vectors of each atom $H^l=(h^l_1,\ldots, h^l_n)$. As the initial approximation of the vector representation of atoms, the trainable vectors corresponding to the charges of the atoms $E: z_i \rightarrow h^0_i$. Since the charges are integers, the mapping $E$ is implemented through a table.

Based on the positions of the atoms, we calculate the matrix of mutual inter atomic distance (IAD) $\text{IAD}_{ij} = \|r_i - r_j\|$. This matrix is invariant with respect to rotation and permutation of the atoms. The one-dimensional value of the distance between atoms must be embedded in a high-dimensional space. It can be realised in many ways \cite{schutt2017schnet, klicpera_dimenet_2020}, from which we choose a simple trigonometric expansion with a uniform frequency step. Radius base function (RBF) is described by the formula
\begin{equation}
    \text{RBF}(r) = \left[\sin(f_0 r/R),\ldots, \sin(f_{K-1} r/R), \cos(f_0 r/R),\ldots, \cos(f_{K-1} r/R)\right]
\end{equation}
where $f_k = f_{\min}\left(\frac{f_{\max}}{f_{\min}}\right)^{k/(K-1)}$, $f_{\min}$ - minimum frequency, $f_{\max}$ - maximum frequency, $K$ - number of periods, $R_{rbf}$ - distance scaling parameter.

The multi-head convolution block implements $H$ graph convolution blocks in parallel. Differentiable distance cutoff mask is calculated from the IAD matrix
\begin{equation}
    \text{DDCM}_{ij} = 1 - \sigma\left(S_{cut} \cdot (\text{IAD}_{ij} - R_{cut})\right)
\end{equation}
where $S_{cut}$ is a transition slope, $R_{cut}$ is a threshold distance.
The convolution block is described by the formula
\begin{equation}
    e_i = \sum_j \left(fr_{ij}\cdot DDCM_{ij}\right)\odot v_j
\end{equation}
Here $v_i$ is the projection of the vector representation of the atom, $fr_{ij}$ is the vector encoding the distance between the atoms $i$ and $j$ (see figure~\ref{fig:convolution_model_scheme}), $\odot$ - element-wise multiplication.

As an activation function, the model uses LeakySiLU, a differentiable analogue of LeakyReLU, defined by the formula
\begin{equation}
    \label{leakySiLU}
    \text{LeakySiLU}(x) = x\left[(1 - ns)\sigma(s x) + ns\right]
\end{equation}
here $\sigma(x)$ is the sigma function, $ns$ is the slope of the straight line at $x \rightarrow -\infty$, $s$ is the slope of transition from the negative to the positive axis.

The energy of the structure is represented as the sum of the average energy of the atom $E_0(z)$ and the correction $\delta E(Z)$ arising due to the interaction of atoms
\begin{equation}
    E = \sum_{z\in Z}E_0(z) + \delta E(Z)
\end{equation}
The values of $E_0(z)$ are calculated using linear regression assuming that $\delta E(Z)$ is absent. The task of the model is to calculate the correction $\delta E(Z)$.

\newpage

\section{Other found cluster structures}
Examples of the clusters found using the artificial potential model trained on $Si_{n}O_{2n\pm1}$, $n=10, \ldots, 30$
\begin{figure}[H]
    \centering
    \begin{subfigure}{0.13\textwidth}
        \centering
        \includegraphics[width=\textwidth]{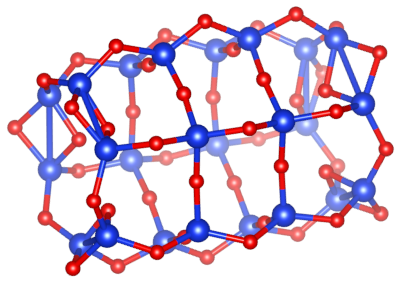}
        \caption*{$Si_{24} O_{48}$}
    \end{subfigure}
    \begin{subfigure}{0.2\textwidth}
        \centering
        \includegraphics[width=\textwidth]{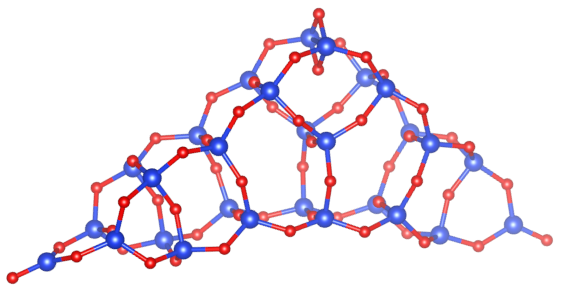}
        \caption*{$Si_{28} O_{56}$}
    \end{subfigure}
    \begin{subfigure}{0.2\textwidth}
        \centering
        \includegraphics[width=\textwidth]{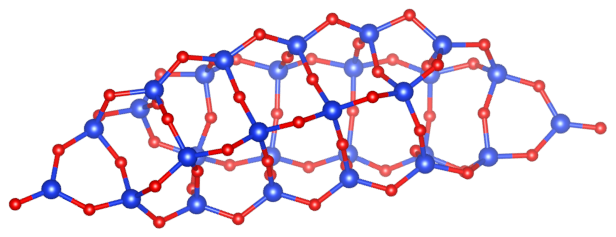}
        \caption*{$Si_{28} O_{56}$}
    \end{subfigure}
    \begin{subfigure}{0.15\textwidth}
        \centering
        \includegraphics[width=\textwidth]{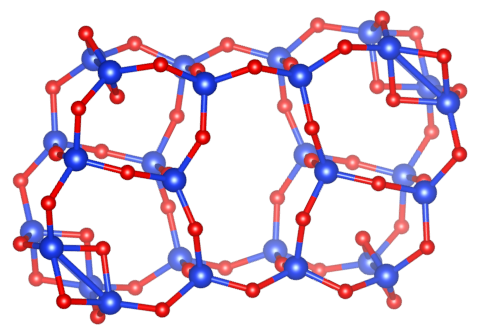}
        \caption*{$Si_{28} O_{56}$}
    \end{subfigure}
    \begin{subfigure}{0.2\textwidth}
        \centering
        \includegraphics[width=\textwidth]{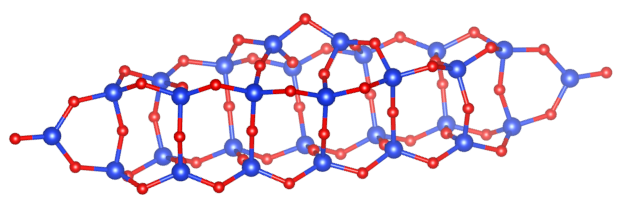}
        \caption*{$Si_{28} O_{56}$}
    \end{subfigure}
    \begin{subfigure}{0.2\textwidth}
        \centering
        \includegraphics[width=\textwidth]{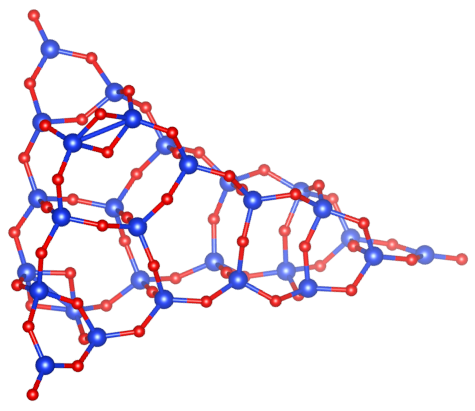}
        \caption*{$Si_{29} O_{58}$}
    \end{subfigure}
    \begin{subfigure}{0.17\textwidth}
        \centering
        \includegraphics[width=\textwidth]{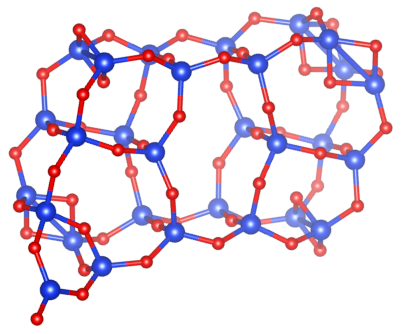}
        \caption*{$Si_{29} O_{58}$}
    \end{subfigure}
    \begin{subfigure}{0.2\textwidth}
        \centering
        \includegraphics[width=\textwidth]{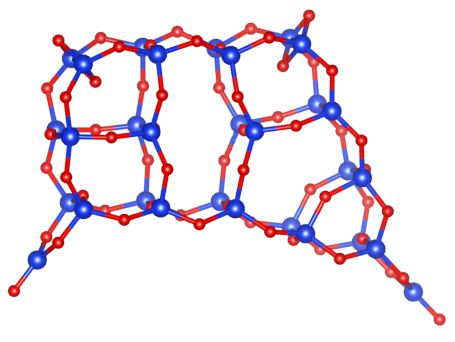}
        \caption*{$Si_{30} O_{60}$}
    \end{subfigure}
    \begin{subfigure}{0.18\textwidth}
        \centering
        \includegraphics[width=\textwidth]{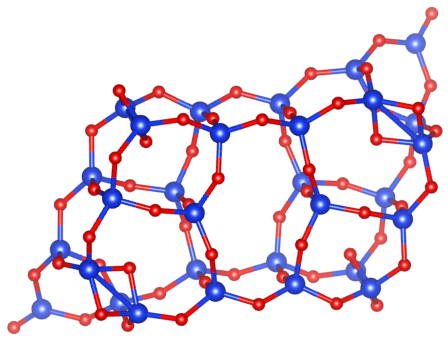}
        \caption*{$Si_{30} O_{60}$}
    \end{subfigure}
    \begin{subfigure}{0.20\textwidth}
        \centering
        \includegraphics[width=\textwidth]{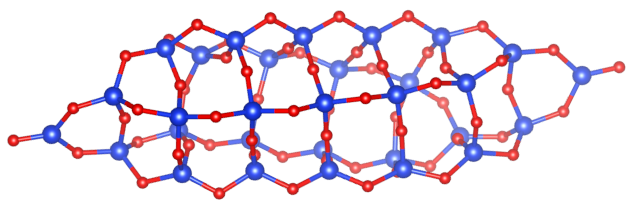}
        \caption*{$Si_{30} O_{60}$}
    \end{subfigure}
    \begin{subfigure}{0.18\textwidth}
        \centering
        \includegraphics[width=\textwidth]{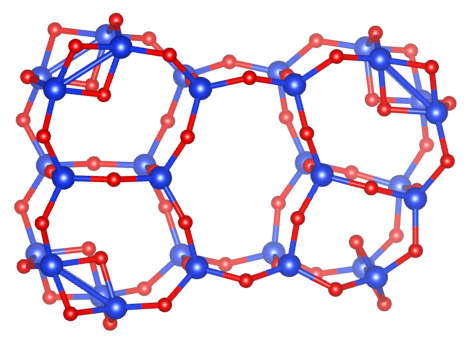}
        \caption*{$Si_{30} O_{60}$}
    \end{subfigure}
    \begin{subfigure}{0.2\textwidth}
        \centering
        \includegraphics[width=\textwidth]{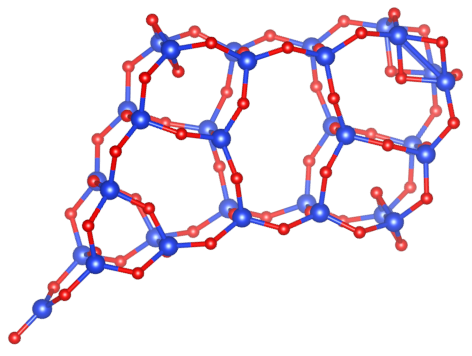}
        \caption*{$Si_{31} O_{62}$}
    \end{subfigure}
    \begin{subfigure}{0.15\textwidth}
        \centering
        \includegraphics[width=\textwidth]{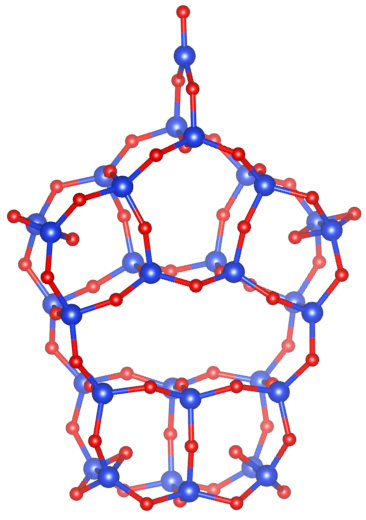}
        \caption*{$Si_{31} O_{62}$}
    \end{subfigure}
    \begin{subfigure}{0.15\textwidth}
        \centering
        \includegraphics[width=\textwidth]{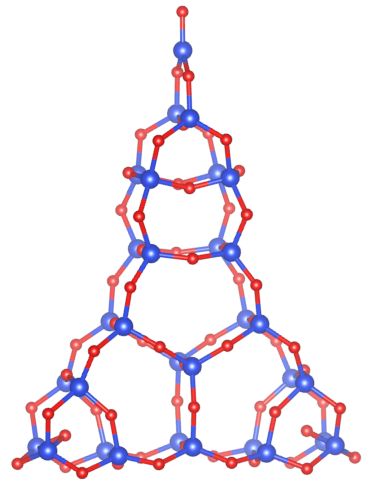}
        \caption*{$Si_{32} O_{64}$}
    \end{subfigure}
    \begin{subfigure}{0.17\textwidth}
        \centering
        \includegraphics[width=\textwidth]{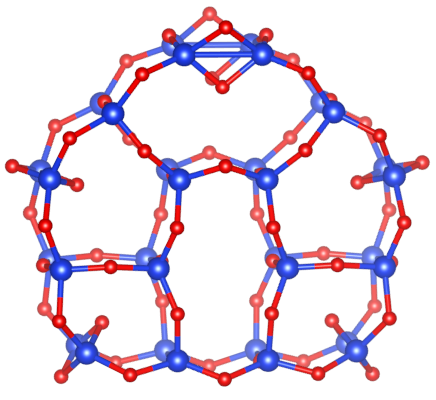}
        \caption*{$Si_{32} O_{64}$}
    \end{subfigure}
    \begin{subfigure}{0.12\textwidth}
        \centering
        \includegraphics[width=\textwidth]{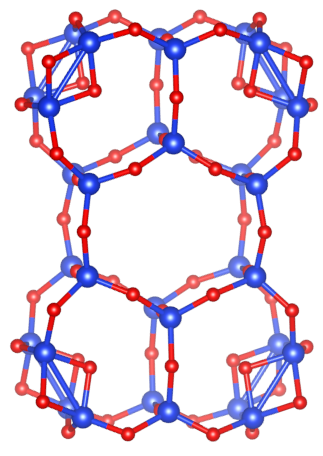}
        \caption*{$Si_{32} O_{64}$}
    \end{subfigure}
    \begin{subfigure}{0.2\textwidth}
        \centering
        \includegraphics[width=\textwidth]{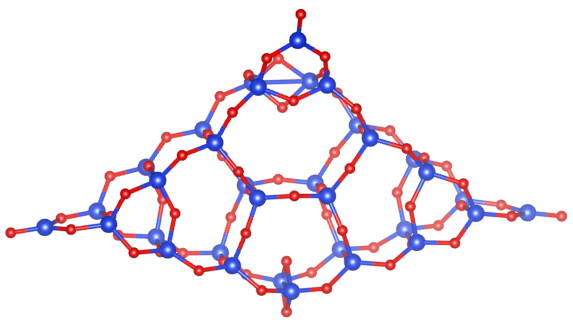}
        \caption*{$Si_{33} O_{66}$}
    \end{subfigure}
    \begin{subfigure}{0.2\textwidth}
        \centering
        \includegraphics[width=\textwidth]{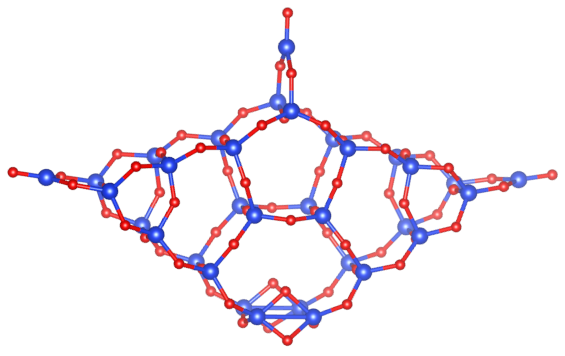}
        \caption*{$Si_{33} O_{66}$}
    \end{subfigure}
    \begin{subfigure}{0.20\textwidth}
        \centering
        \includegraphics[width=\textwidth]{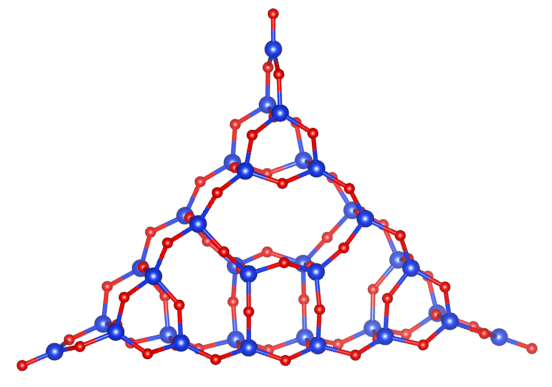}
        \caption*{$Si_{33} O_{66}$}
    \end{subfigure}
    \begin{subfigure}{0.14\textwidth}
        \centering
        \includegraphics[width=\textwidth]{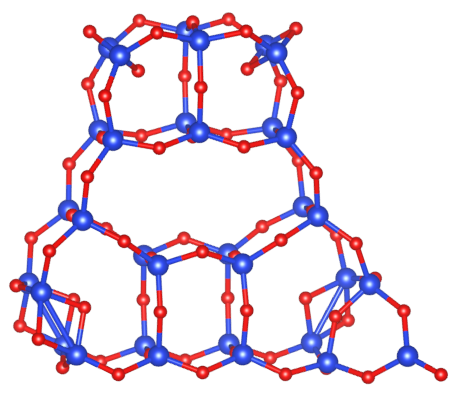}
        \caption*{$Si_{33} O_{66}$}
    \end{subfigure}
    \begin{subfigure}{0.2\textwidth}
        \centering
        \includegraphics[width=\textwidth]{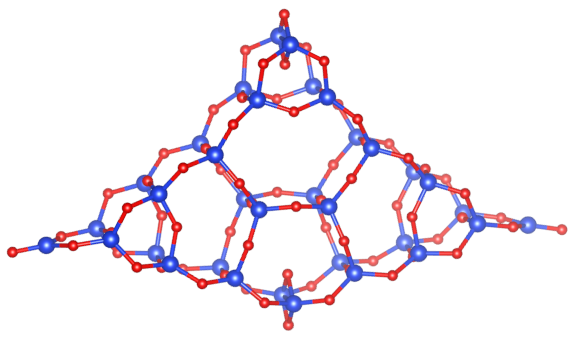}
        \caption*{$Si_{34} O_{68}$}
    \end{subfigure}
    \begin{subfigure}{0.2\textwidth}
        \centering
        \includegraphics[width=\textwidth]{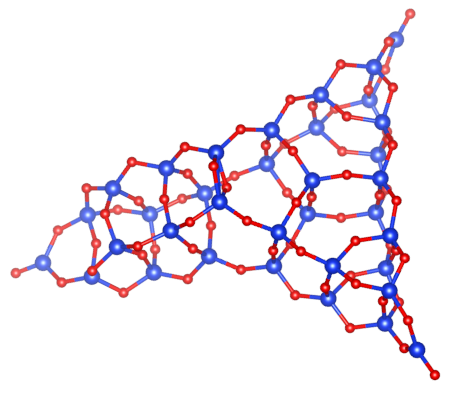}
        \caption*{$Si_{35} O_{70}$}
    \end{subfigure}
    \begin{subfigure}{0.20\textwidth}
        \centering
        \includegraphics[width=\textwidth]{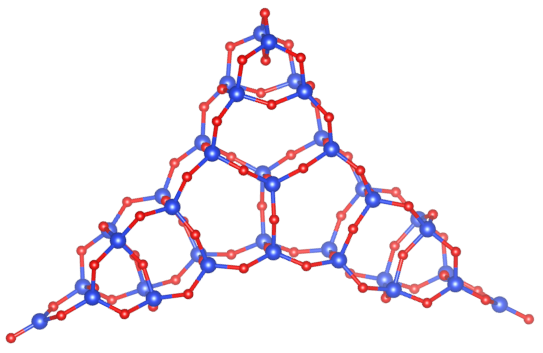}
        \caption*{$Si_{36} O_{72}$}
    \end{subfigure}
    \begin{subfigure}{0.18\textwidth}
        \centering
        \includegraphics[width=\textwidth]{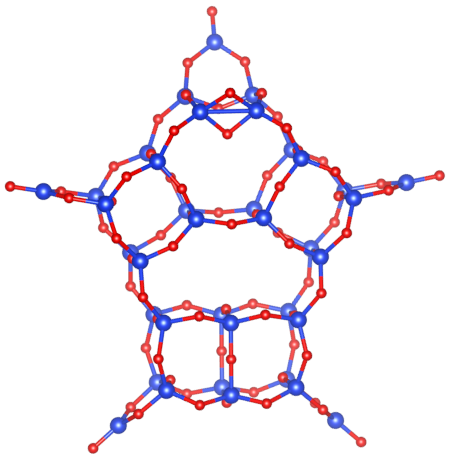}
        \caption*{$Si_{37} O_{74}$}
    \end{subfigure}
    \begin{subfigure}{0.16\textwidth}
        \centering
        \includegraphics[width=\textwidth]{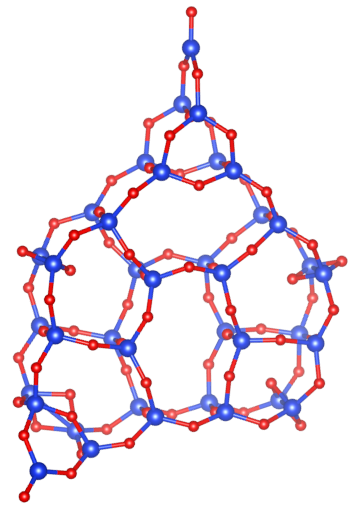}
        \caption*{$Si_{38} O_{76}$}
    \end{subfigure}
    \begin{subfigure}{0.16\textwidth}
        \centering
        \includegraphics[width=\textwidth]{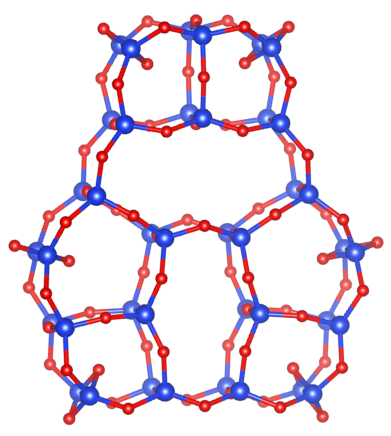}
        \caption*{$Si_{40} O_{80}$}
    \end{subfigure}
    \begin{subfigure}{0.18\textwidth}
        \centering
        \includegraphics[width=\textwidth]{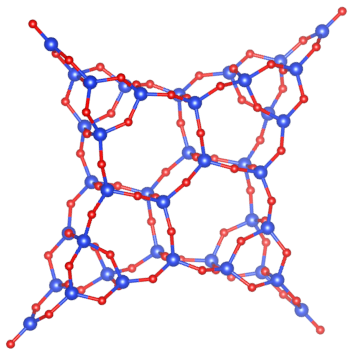}
        \caption*{$Si_{40} O_{80}$}
    \end{subfigure}
    \begin{subfigure}{0.2\textwidth}
        \centering
        \includegraphics[width=\textwidth]{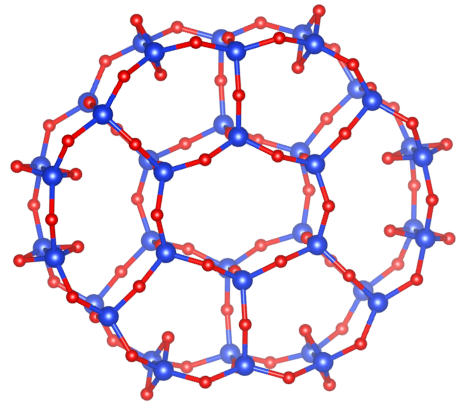}
        \caption*{$Si_{40} O_{80}$}
    \end{subfigure}
    \begin{subfigure}{0.17\textwidth}
        \centering
        \includegraphics[width=\textwidth]{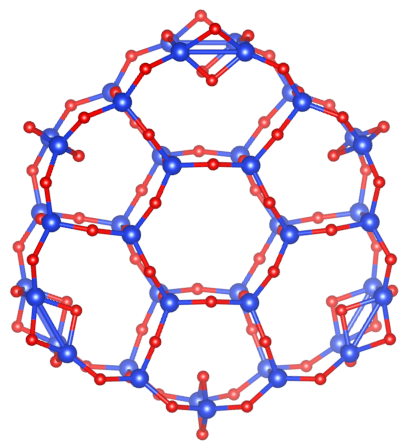}
        \caption*{$Si_{42} O_{84}$}
    \end{subfigure}
    \begin{subfigure}{0.17\textwidth}
        \centering
        \includegraphics[width=\textwidth]{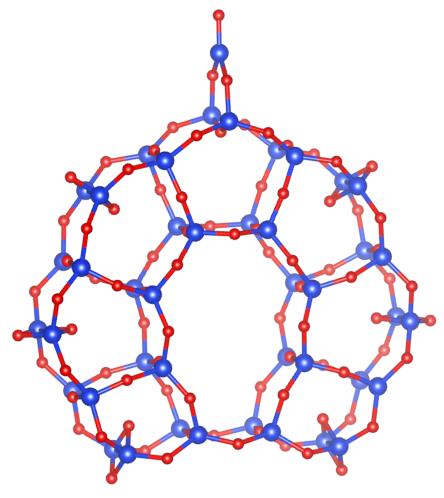}
        \caption*{$Si_{43} O_{86}$}
    \end{subfigure}
\end{figure}

\begin{figure}[H]
    \centering
    \begin{subfigure}{0.18\textwidth}
        \centering
        \includegraphics[width=\textwidth]{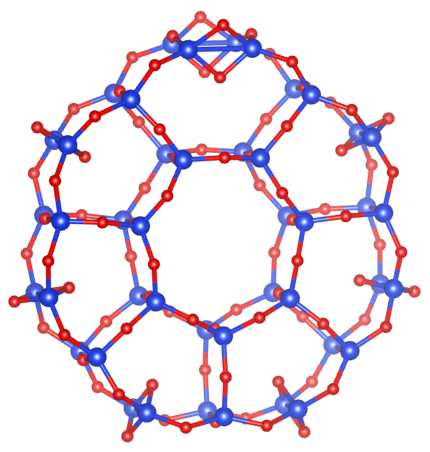}
        \caption*{$Si_{44} O_{88}$}
    \end{subfigure}
    \begin{subfigure}{0.19\textwidth}
        \centering
        \includegraphics[width=\textwidth]{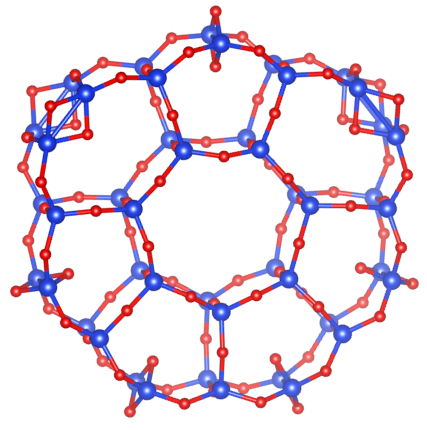}
        \caption*{$Si_{46} O_{92}$}
    \end{subfigure}
    \begin{subfigure}{0.2\textwidth}
        \centering
        \includegraphics[width=\textwidth]{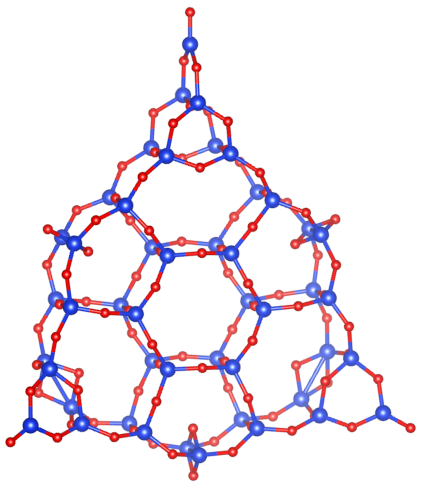}
        \caption*{$Si_{47} O_{94}$}
    \end{subfigure}
    \begin{subfigure}{0.2\textwidth}
        \centering
        \includegraphics[width=\textwidth]{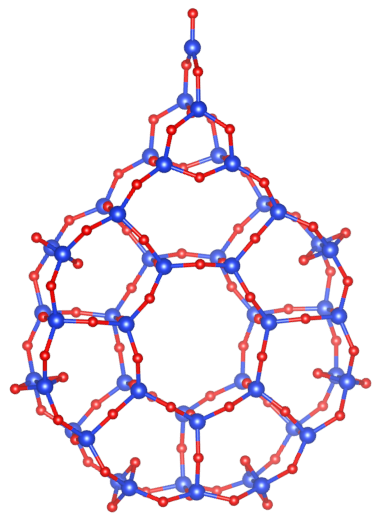}
        \caption*{$Si_{47} O_{94}$}
    \end{subfigure}
    \begin{subfigure}{0.20\textwidth}
        \centering
        \includegraphics[width=\textwidth]{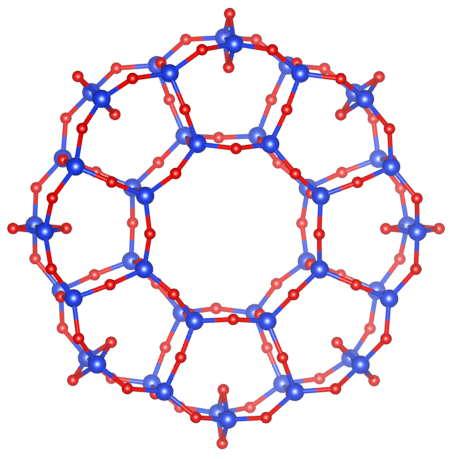}
        \caption*{$Si_{48} O_{96}$}
    \end{subfigure}
    \begin{subfigure}{0.2\textwidth}
        \centering
        \includegraphics[width=\textwidth]{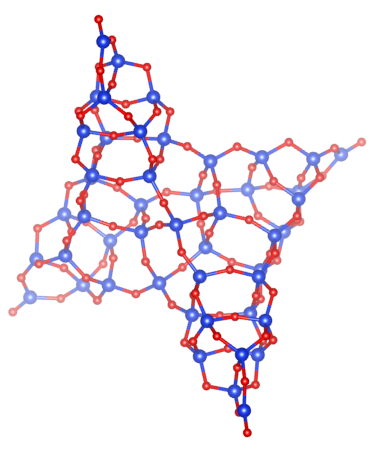}
        \caption*{$Si_{48} O_{96}$}
    \end{subfigure}
\end{figure}

\end{document}